\documentclass[11pt,letterpaper]{article}
\usepackage{amsmath,amssymb}    % AMS
\usepackage{color}
\usepackage{algorithm}
\usepackage{algorithmic}
\usepackage{graphicx,float}     % figures
\usepackage{caption}
\usepackage{subcaption}
\usepackage{vmargin}   % margins ,fancyhdr

\setpapersize{USletter}
\setmarginsrb{.75in}{.5in}        % left, top
             {.75in}{.5in}        % right, bottom
             {.25in}{.25in}     % headheight, headsep
             {.25in}{.5in}      % footheight, footskip
\floatstyle{ruled}
\newfloat{algo}{tbp}{lop}%[section]
\floatname{algo}{Algorithm}

\newtheorem{theorem}{Theorem}[section]

\newtheorem{corollary}[theorem]{Corollary}
\newtheorem{definition}[theorem]{Definition}

\newtheorem{lemma}[theorem]{Lemma}
\newtheorem{fact}[theorem]{Fact}

\newcommand{\Proof}[0]{\smallskip\noindent\textit{\textbf{Proof}}\quad}
\newcommand{\Proofof}[1]{\smallskip\noindent\textit{\textbf{Proof of #1:}}\quad}
\newcommand{\QED}[0]{\hfill\ensuremath{\blacksquare}\medspace}

\newcommand{\expdistr}{\textit{Exp}}
\newcommand{\dist}{\text{dist}}
\newcommand{\distshift}{\text{dist}_{- \delta}}

\newcommand{\prob}[2][\mbox{}]{\ensuremath{\mathop{\text{\normalfont {\textbf{Pr}}}}_{#1}\left[#2\right]}}
\newcommand{\expct}[2][\mbox{}]{\ensuremath{\mathop{\text{\normalfont \textbf{E}}}_{#1}}\left[#2\right]}

\begin{document}

\title{Parallel Graph Decompositions Using Random Shifts
\thanks{Partially supported by the National Science Foundation
under grant number CCF-1018463 and CCF-1065106}
}
%\numberofauthors{3}
\author{
  Gary L.\ Miller\\
  CMU\\
  \texttt{glmiller@cs.cmu.edu}\\
  \and
  Richard Peng \thanks{Supported by a Microsoft Fellowship}\\
  CMU\\
  \texttt{yangp@cs.cmu.edu}
  \and
  Shen Chen Xu\\
  CMU\\
  \texttt{shenchex@cs.cmu.edu}
}

\maketitle

\begin{abstract}
We show an improved parallel algorithm for decomposing
an undirected unweighted graph into small diameter pieces
with a small fraction of the edges in between.
These decompositions form critical subroutines in a number
of graph algorithms.
Our algorithm builds upon the shifted shortest path approach
introduced in [Blelloch,Gupta,Koutis,Miller,Peng,Tangwongsan,~SPAA~2011].
By combining various stages of the previous algorithm,
we obtain a significantly simpler algorithm with the same
asymptotic guarantees as the best sequential algorithm.
\end{abstract}

\label{sec:intro}

\section{Introduction}
Graph decomposition aims to partition the vertices of a graph into
well connected pieces so that few edges are between pieces.
A variety of measures of the connectivity within a piece, such as
diameter, conductance, and spectral properties have been studied.
The more intricate measures such as conductance have proven
to be particularly useful in applications \cite{ShiM97},
and have been well studied \cite{LeightonR99, Sherman09}.
However, these algorithms, as well as many others, use simpler
low diameter decompositions as a subroutine.
This variant takes a much more simplistic view of the connectivity
within each piece, and measures it using only the diameter.

The various shortcomings of such decompositions as a stand-alone
routine are offset by its potential as an algorithmic tool.
Low diameter decompositions/clusterings were first introduced
in \cite{Awerbuch85,AwerbuchGLP89}, and have
been used as core subroutine for a number of algorithms such as:
approximations to sparsest cut \cite{LeightonR99,Sherman09};
construction of spanners \cite{Cohen98};
parallel approximations of shortest path in undirected graphs \cite{Cohen00};
and generating low-stretch embedding of graphs into trees
\cite{AlonKPW95,FakcharoenpholRT03,ElkinEST08,AbrahamN12}.

More recently, the connection of low diameter decomposition with generating low
stretch spanning trees was used in \cite{BlellochGKMPT11} to give
nearly-linear work parallel solvers for SDD linear systems.
These solvers can in turn be used as a black-box in algorithms for computing
maximum flow and negative-length shortest path \cite{ChristianoKMST10, DaitchS08}.
This led to parallel algorithms whose work is within polylog factors of
the best known sequential algorithms.
As these are problems for which work-efficient parallelizations have proven
to be elusive, parallel solvers for SDD linear systems represent
a promising new direction for designing parallel algorithms.

The parallel SDD linear system solver algorithm from~\cite{BlellochGKMPT11}
is of mostly theoretical interest due to a large polylog factor in work.
Much of this is due to the nearly-linear work, parallel low diameter
decomposition algorithm introduced in the same paper, which was in
turn used as a subroutine to generate tree embeddings.
Therefore, finding improved parallel graph decomposition routines
represent a natural direction for finding faster parallel solver algorithms.

In order to formally specify the diameter of a piece, it is crucial to
emphasize the distinction between weak and strong diameter.
The diameter of a piece $S \subseteq V$ can be defined in
two ways, weak and strong diameter.
Both of them define diameter to the maximum length of a shortest path between
two vertices in $S$, while the difference is in the set of allowed paths.
{\bf Strong diameter} restricts the shortest path between two vertices in $S$
to only use vertices in $S$, while {\bf weak diameter} allows
for shortcuts through vertices in $V \setminus S$.
The optimal tree metric embedding algorithm \cite{FakcharoenpholRT03}
relies on weak diameter.
It has been parallelized with polylog work overhead \cite{BlellochGT12},
but takes quadratic work.

A trend in algorithms that use weak diameter is that their running time
tends to be quadratic.
This is also the case with parallel algorithms for computing
low diameter decompositions \cite{AwerbuchBCP92}.
To date, nearly-linear work algorithms for finding tree embedding
use strong diameter instead.
While this leads to more difficulties in bounding diameters, the overall work
is easier to bound since each piece certifies its own diameter, and therefore
does not need to examine other pieces.
For SDD linear system solvers \cite{BlellochGT12}, strong
diameter is also crucial because the final tree (which the graph embeds into)
is formed by combining the shortest path tree in each of the pieces.
As a result, we will use diameter of a piece to denote {\bf strong diameter}
for the rest of this paper, and define a low diameter decomposition as
follows:
\begin{definition}
\label{def:lowdiam}
Given an undirected, unweighted graph $G = (V, E)$,
a $(\beta, d)$ decomposition is a partition of $V$ into subsets
$S_1 \ldots S_k$ such that:
	\begin{itemize}
		\item The (strong) diameter of each $S_i$ is at most $d$.
		\item The number of edges with endpoints
			belonging to different pieces is at most $\beta m$.
	\end{itemize}
\end{definition}
A standard choice of parameters for such decompositions is
$(\beta, O(\frac{\log{n}}{\beta} ))$, which are in some sense optimal.
Furthermore, when computing tree embedding, $\beta$ is often set to
$\log^{-c}n$ for some constant $c$.
As a result, the algorithm given in \cite{BlellochGKMPT11}, as well as
our algorithm are geared towards small diameters.
This makes the running time of these algorithms more than the
$\mathcal{NC}$ algorithms such as \cite{AwerbuchBCP92}
for large values of $\beta$.
However, they suffice for the purpose of generating tree/subgraph
embedding in polylog depth, as well as low work parallel graph
algorithms.

Obtaining these  decompositions in the sequential
setting can be done via a process known as ball growing.
This process starts with a single vertex, and repeatedly adds the
neighbors of the current set into the set.
It terminates when the number of edges on the boundary is less
than a $\beta$ fraction of the edges within, which is equivalent
to the piece having small conductance.
Once the first piece is found, the algorithm discards its vertices
and repeats on the remaining graph.
The final bound of $\beta m$ edges between the pieces
can in turn be obtained by summing this bound over all the pieces.
Using a consumption argument, one could prove that the diameter of
a piece does not exceed $O(\frac{\log{n}}{\beta})$.
Because we are okay with a depth that depends on $1 / \beta$,
and the piece's diameters can be bounded by $O(\frac{\log{n}}{\beta})$,
finding a single piece is easy to parallelize.
However, the strong diameter requirement means that we cannot
start finding the second ball until we are done finding the first.
This leads to a chain of sequential dependencies that may be
as long as $\Omega(n)$, and is the main challenge in
obtaining a parallel decomposition algorithm.

The parallel SDD linear system solver algorithm given in
\cite{BlellochGKMPT11} relied on a parallel algorithm
that computes a $O(\beta, \frac{\log^4{n}}{\beta})$ decomposition
in $O(\frac{\log^3{n}}{\beta})$ depth and $O(m\log^2{n})$ work.
This algorithm showed that some of the ball growing steps
can be performed simultaneously in parallel,
leading to balls which have small overlap.
Then a randomly shifted shortest path routine is used to
resolve these overlaps.
In this paper we show that these two steps can be combined
in a simple, global routine.
This leads to a simple algorithm that picks random shifts
in a more intricate way and assigns vertices to pieces using
one shortest path invocation.
In the PRAM model, our result can be described by the following theorem:
\begin{theorem}
\label{thm:partition}
	There is an algorithm \textsc{Partition} that takes an unweighted graph
with $n$ vertices, $m$ edges, a  parameter $\beta \leq 1/2$
and produces a $(\beta, O(\frac{\log{n}}{\beta}))$ decomposition
in expected $O(\frac{\log^2{n}}{\beta})$ depth and $O(m)$ work.
\end{theorem}

We will give an overview of our algorithm in Section \ref{sec:related} and
define our notations and give relevant background in Section
\ref{sec:notations}. Section \ref{sec:partition} contains the analysis of our
partition routine, and modifications to make it more suitable for
implementation and parallelization are given in Section \ref{sec:implementation}.
In Section \ref{sec:conclusion} we discuss some possible
extensions to our algorithm.

\section{Overview and Related Works}
\label{sec:related}

In this section we give an intuitive view of our partition routine and discuss
how it relates to various other graph decomposition algorithms that have been
studied in the past. We will defer the implementation details to Section
\ref{sec:implementation}. A simple interpretation of our algorithm executing in
parallel is outlined in Algorithm \ref{alg:parallelPartition}.

\begin{algo}[ht]
\textsc{Parallel Partition}
\vspace{0.05cm}

\underline{Input:}
Undirected, unweighted graph $G = (V, E)$, parameter $0 < \beta < 1$ and
parameter $d$ indicating failure probability.

\underline{Output:} $(\beta, O(\log{n} / \beta))$ decomposition of $G$ with
probability at least $1 - n^{-d}$.

\begin{algorithmic}[1]
	\STATE{\textit{IN PARALLEL} each vertex $u$  picks $\delta_u$
       independently from an exponential distribution with mean $1/\beta$.}
        \STATE{\textit{IN PARALLEL} compute $\delta_{\max} = \max\{ \delta_u \; | \; u
          \in V\}$}
	\STATE{Perform \textit{PARALLEL BFS}, with vertex $u$ starting when the vertex at the head of the queue has distance more than $\delta_{\max} - \delta_u$.}
       \STATE{\textit{IN PARALLEL} Assign each vertex $u$ to point of origin of the shortest path that reached it in the BFS.}
\end{algorithmic}

\caption{Parallel Partition Algorithm}
\label{alg:parallelPartition}

\end{algo}

\begin{figure*}[ht]
\centering
    \begin{subfigure}{0.32\linewidth}
      \includegraphics[width=\textwidth]{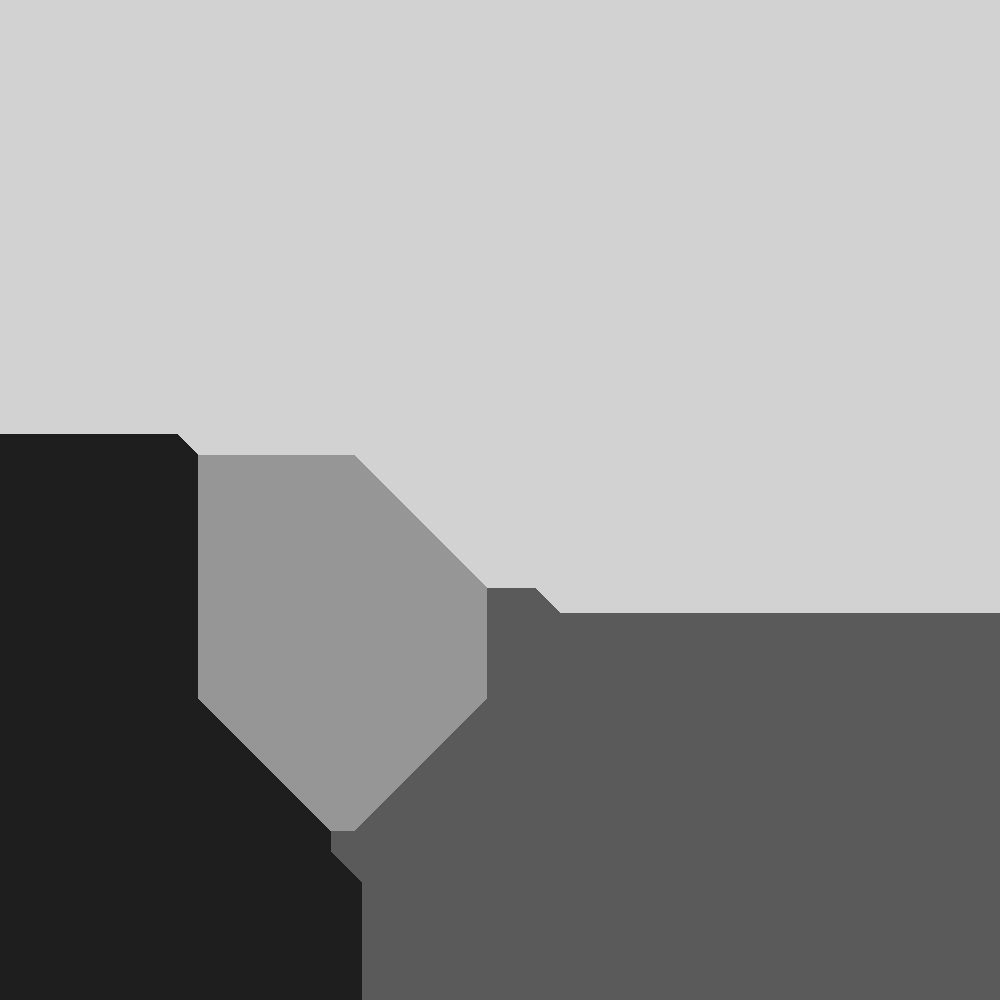}
      \caption{$\beta=0.002$}
    \end{subfigure}
    \begin{subfigure}{0.32\linewidth}
      \includegraphics[width=\textwidth]{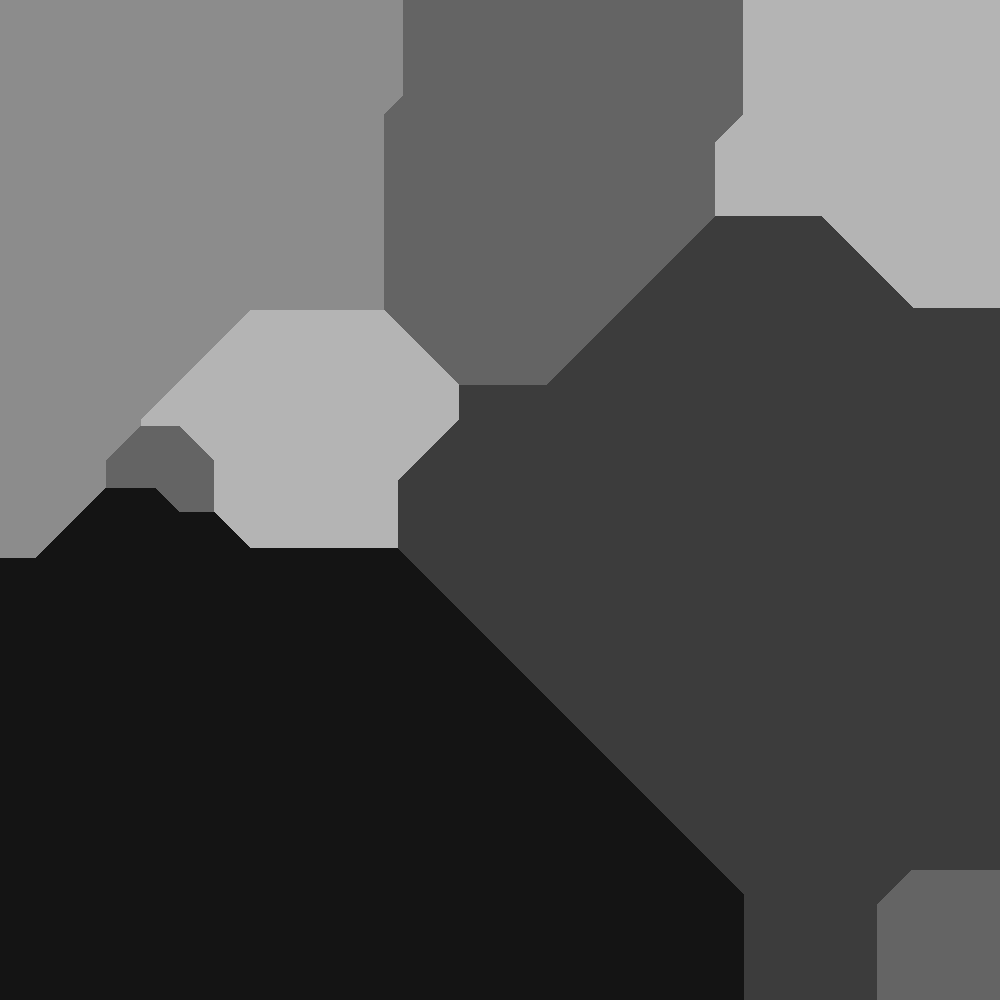}
      \caption{$\beta=0.005$}
    \end{subfigure} 
   \begin{subfigure}{0.32\linewidth}
      \includegraphics[width=\textwidth]{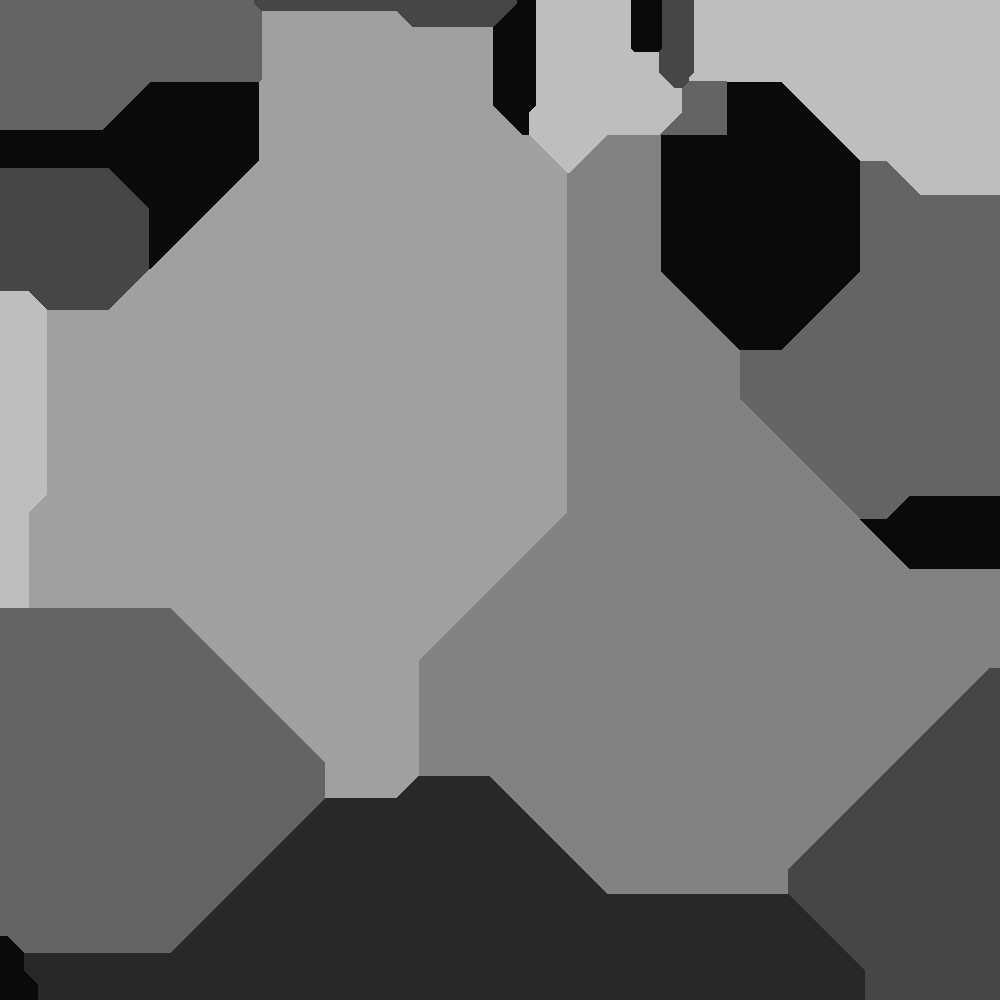}
      \caption{$\beta=0.01$}
    \end{subfigure}
    \\
    \begin{subfigure}{0.32\linewidth}
      \includegraphics[width=\textwidth]{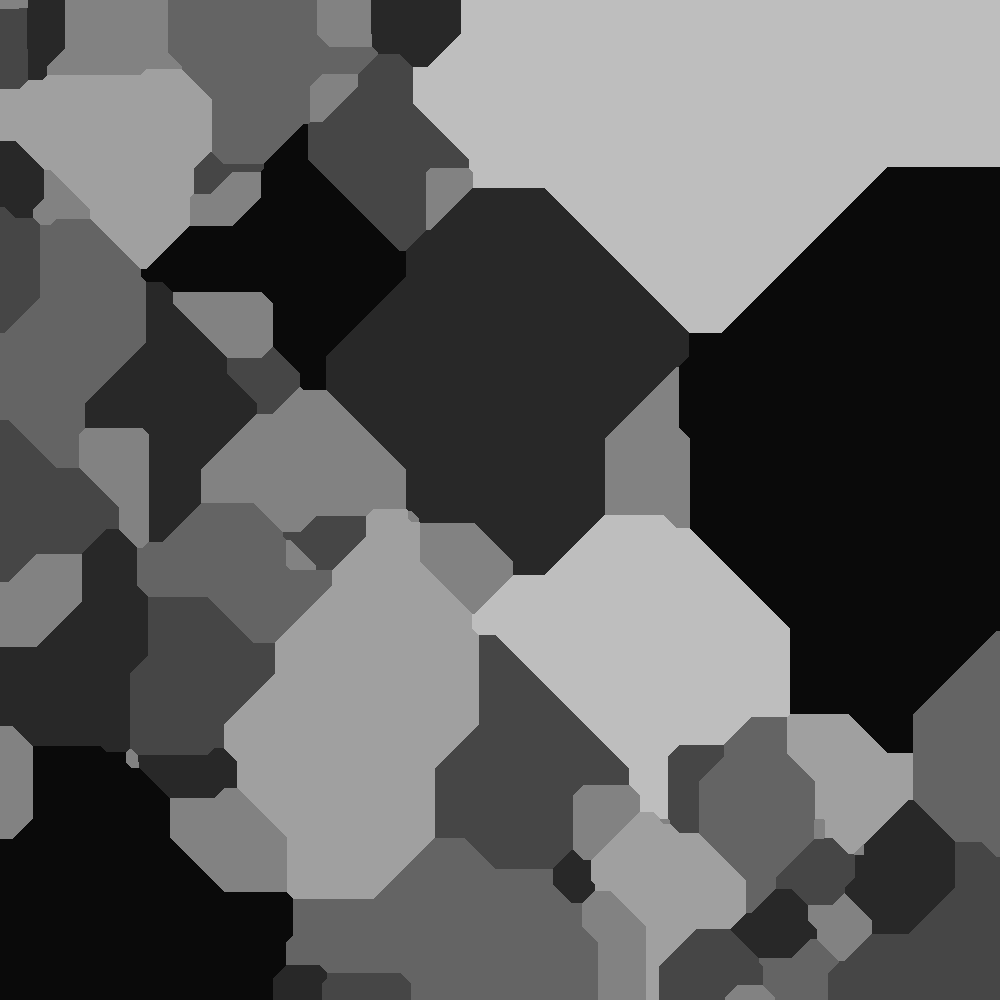}
      \caption{$\beta=0.02$}
    \end{subfigure}
    \begin{subfigure}{0.32\linewidth}
      \includegraphics[width=\textwidth]{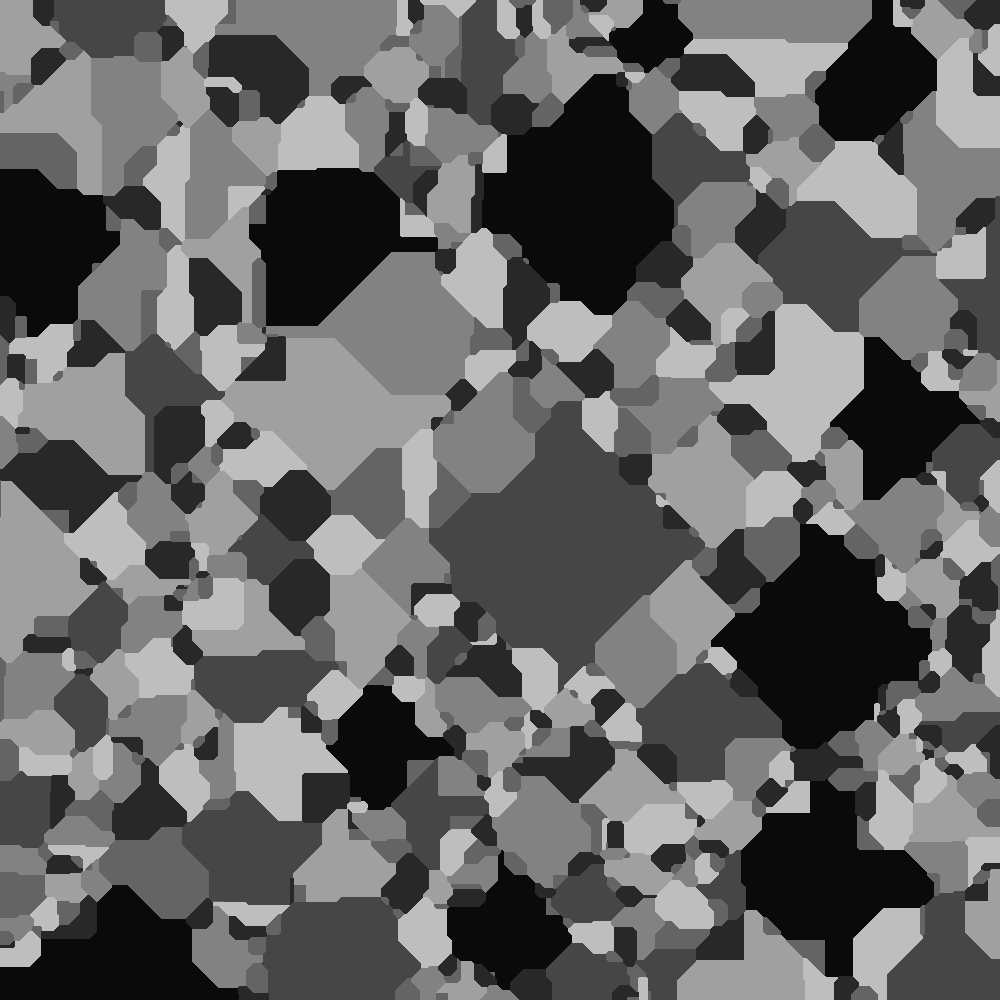}
      \caption{$\beta=0.05$}
    \end{subfigure}
    \begin{subfigure}{0.32\linewidth}
     \includegraphics[width=\textwidth]{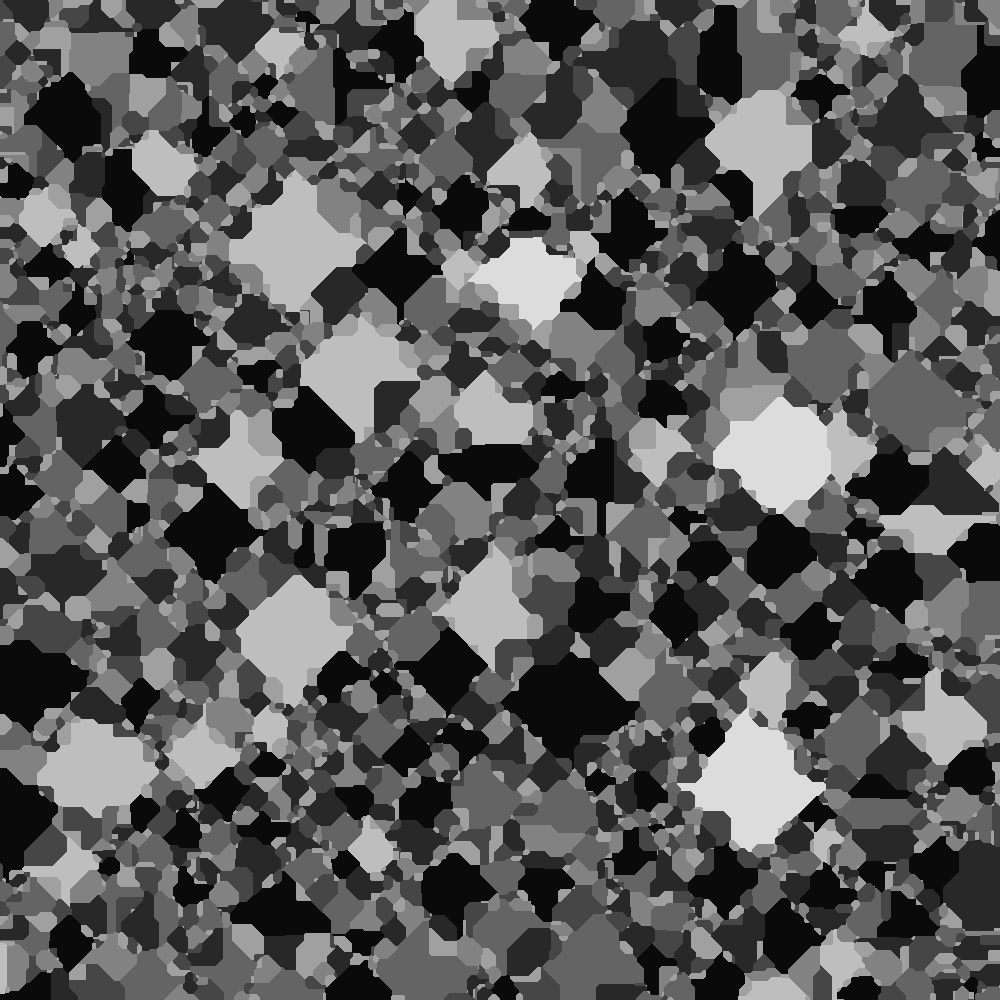}
      \caption{$\beta=0.1$}
    \end{subfigure}
  \caption{Decompositions generated by our algorithm on a
  $1000 \times 1000$ grid under varying values of $\beta$.
  Different colors represent different clusters}
  \label{fig:squareGrid}
\end{figure*}

Steps 1, 2, and 4 of our algorithm are done independently at each vertex
and are clearly parallelizable.
So the main algorithmic aspects of our algorithm is in step 3, which
is performing a breadth first search while recording the point of origin.
Such processes have been well-studied~\cite{LeisersonS10, BeamerAP12},
and we will discuss how to use such routines in a black-box manner
in Section~\ref{sec:implementation}.
More intuitively, our algorithm can also be viewed as performing
parallel ball growing with random delays.
Each vertex $u$ picks a start time according to some distribution, and
if $u$ is not already part of some other cluster at that time, $u$ starts a
cluster of its own and performs a breadth first search.
The search takes one unit of time to propagate across an edge,
and each such time steps can be performed in parallel over all vertices.
If a vertex $v$ visited during the search is not yet part of
any other cluster, it joins the cluster of the vertex that first reached it,
and its neighbors are added to the BFS queue.
%If $v$ is already owned by some other cluster when it is
%reached, we do not search its neighbors.
The randomized start times leads to both required properties of the
decompositions, as well as small bound on the depth of the BFS trees.
Figure \ref{fig:squareGrid} shows the resulting partitions of our algorithm
with a $1000 \times 1000$ square grid as input and different values
of $\beta$ used to generate the delays.
Lower $\beta$ leads to larger diameter and fewer edges on the boundaries,
which matches our more detailed analysis in Section~\ref{sec:partition}.

%%
%%\begin{figure}[h]
%%  \begin{center}
%%    \begin{subfigure}{0.49\linewidth}
%%      \includegraphics[width=\textwidth]{grid-0_002.png}
%%      \caption{$\beta=0.002$}
%%    \end{subfigure}
%%    \begin{subfigure}{0.49\linewidth}
%%      \includegraphics[width=\textwidth]{grid-0_005.png}
%%      \caption{$\beta=0.005$}
%%    \end{subfigure}
%%    \\
%%    \begin{subfigure}{0.49\linewidth}
%%      \includegraphics[width=\textwidth]{grid-0_01.png}
%%      \caption{$\beta=0.01$}
%%    \end{subfigure}
%%    \begin{subfigure}{0.49\linewidth}
%%      \includegraphics[width=\textwidth]{grid-0_02.png}
%%      \caption{$\beta=0.02$}
%%    \end{subfigure}
%%    \\
%%    \begin{subfigure}{0.49\linewidth}
%%      \includegraphics[width=\textwidth]{grid-0_05.png}
%%      \caption{$\beta=0.05$}
%%    \end{subfigure}
%%    \begin{subfigure}{0.49\linewidth}
%%      \includegraphics[width=\textwidth]{grid-0_1.png}
%%      \caption{$\beta=0.1$}
%%    \end{subfigure}
%%  \end{center}
%%  \caption{Decompositions generated by our algorithm on a
%%  $1000 \times 1000$ grid under varying values of $\beta$.
%%  Different colors represent different clusters}
%%  \label{fig:squareGrid}
%%\end{figure}

To our knowledge, low diameter decompositions as stated in
Definition~\ref{def:lowdiam} were first
used for distributed algorithms in~\cite{Awerbuch85}.
Subsequently it has been used as a key subroutine in the construction of
low-stretch spanning trees algorithms~\cite{AlonKPW95}, or more
generally embeddings of graphs into trees~\cite{Bartal96}.
Another application of unweighted decompositions is for efficiently
computing separators in minor-free graphs~\cite{PlotkinRS94,Wulffnilsen11}.
Our algorithm can be directly substituted into these algorithms,
although the main improvements that we obtain are for generating
low stretch spanning trees using the framework of~\cite{BlellochGKMPT11}.

A definition related to low diameter decomposition is block
decompositions from~\cite{LinialS91}.
One of their main algorithmic routines is to partition a graph into
$O(\log{n})$ blocks such that each connected piece in a block
has diameter $O(\log{n})$.
This decomposition can also be obtained by iteratively running
a $(\frac{1}{2}, O(\log{n}))$ low diameter decomposition $O(\log{n})$ times.
This is because the number of edges not in a block decreases by
a factor of $2$ per iteration.

The main scheme of our algorithm is to pick radii of the balls independently
from some distribution.
Similar approaches have been used in computing block decompositions
in ~\cite{LinialS91}, as well as finding $(r, \rho, \gamma)$-probabilistic
partitions needed for the Bartal trees~\cite{Bartal96}.
Our partition scheme differs from these in that the process behaves
identically on all vertices, and our guarantees are in terms of strong
diameter.
The first difference means that the formation of clusters in our algorithm
does not have sequential dependencies; while the later actually leads
to a lower work term.
A typical method for meeting weak diameter requirements is to broadcast
to all vertices within a certain distance~\cite{Awerbuch85,LinialS91}.
As the graph may have small diameter, this can lead to work that is
quadratic in the number of vertices.
By broadcasting only along shortest paths in a way that is akin to breadth
first search, we are able to reduce this to $O(m)$ work.

Aside from being directly related to decomposition schemes for
unweighted undirected graphs, our algorithm can also be adapted
to give guarantees similar to other improved decomposition schemes.
When viewed as a sequential algorithm, it can also lead to similar guarantees
on weighted graphs to the decomposition scheme from~\cite{Bartal96}
as well as generalizations needed for improved low stretch spanning
tree algorithms~\cite{Bartal96,ElkinEST08}.
However, the parallel performance of our algorithm in the weighted
setting is less clear, and we will describe this question in more detail
in Section~\ref{sec:conclusion}.

\section{Background and Notations}
\label{sec:notations}

In this section we state some standard notations, and review some
key ideas introduced in the Blelloch et al. algorithm \cite{BlellochGKMPT11}.
Given a graph $G$, we use $\dist(u, v)$ to denote the length of
the shortest path from $u$ to $v$.
As with earlier works on tree embedding,
we will pick a special vertex in each piece, and use the distance to
the farthest vertex from it as an estimate for the diameter.
This simplification can be made since the graph is undirected and
the final bound allows for constant factors (specifically $2$).
We will denote this special vertex the center of the piece, and denote
the piece centered at $u$ using $S_u$.

As the number of pieces in the final decomposition may be large
(e.g. the line graph), a parallel algorithm needs to construct a number of
pieces simultaneously.
On the other hand, for closely connected graphs such as the complete
graph, a single piece may contain the entire graph.
As a result, if too many pieces are grown independently, the total
work may become quadratic.
The decomposition algorithm by Blelloch et al. \cite{BlellochGKMPT11}
addressed this tradeoff by gradually increasing the number of
centers picked iteratively.
It was motivated by the $(\beta, W)$ decompositions used
in an algorithm by Cohen for approximating shortest paths
in undirected graphs \cite{Cohen00}.
By running iterations with gradually more centers,
it can be shown that the resulting pieces at
each iteration have small overlap.
This overlap is in turn resolved using a shifted shortest
path algorithm, which introduces shifts (denoted by $\delta$) at the centers
and assigns vertex $v$ to $S_u$ that minimizes the
shifted distance:
\begin{align}
\distshift(u, v) = \dist(u, v) - \delta_u.
\label{eqn:shiftDist}
\end{align}
It was shown that by picking shifts uniformly from a sufficiently
large range, a $(\beta, O(\frac{\log^{c} n}{\beta}))$ decomposition
can be obtained.

Our algorithm can be viewed as a more streamlined algorithm
that combines these two components.
Note that sampling vertices with exponentially increasing
density can be emulated by adding a large, step-like increase
to the shifts of centers picked in earlier iterations.
Furthermore, the need to have exponentially decreasing number
of centers in the iterations suggests that the exponential
distribution can be used in place of the (locally) uniform distribution.
This distribution has been well-studied, and the properties of it
that we will need have been used to study its order statistics
in fault tolerance \cite{Trivedi02}.
For a parameter $\gamma$, this distribution
is defined by the density function:
\begin{align*}
f_{Exp}(x, \gamma) =
\left\{
\begin{array}{lr}
\gamma \exp(- \gamma x) & \text{if $x \geq 0$}, \\
0 & \text{otherwise}.
\end{array}
\right.
%\label{eq:df}
\end{align*}
We will denote it using $\expdistr(\gamma)$
and will also make use of its cumulative density function:

\begin{align*}
F_{Exp}(x, \gamma) =
\prob{\expdistr(\gamma) \leq x} &=
\left\{
\begin{array}{lr}
1 - \exp(- \gamma x) & \text{if $x \geq 0$}, \\
0 & \text{otherwise}.
\end{array}
\right.
% \label{eq:cdf}
\end{align*}

A crucial fact about the exponential distribution is that it is memoryless.
That is, if we condition on $\expdistr(\gamma) \geq t$,
then $\expdistr(\gamma) - t$ will follow the same distribution.
We will also use order statistic of random independent variables
following the exponential distribution.
Given $n$ random variables $X_1 \ldots X_n$, the $i$\textsuperscript{th}
order statistic of them is  the value of the $i$\textsuperscript{th} smallest random variable.
Another property of exponential distributions is that the difference
between its order statistics also follow exponential distributions.
The following fact as stated on page 19 of~\cite{Feller71}
has been used in a variety of settings.

\begin{fact}
\label{fact:orderstatisticexp}
The $n$ variables $X_{(1)}, X_{(2)}- X_{(1)}, \cdots, X_{(n)} - X_{(n - 1)}$
are independent and the density of $X_{(k + 1)} - X_{(k)}$
is given by $f_{Exp}(x, \gamma)$ where $\gamma = (n - k)$.
\end{fact}

Let $X_{(i)}^n$ denote the $i$\textsuperscript{th} order statistic of $n$
i.i.d.\ exponential random variables. An intuitive way to prove the above fact
is that by the i.i.d assumption, the cumulative density distribution of
$X_{(1)}^n$ is given by
\begin{align*}
  F_{X^n_{(1)}}(x)&=1-(1-F(x))^n
  \\&=1-\exp(-n\beta x)
\end{align*}
where $F(x)=1-\exp(-\beta x)$ is the cumulative distribution of
$\mathit{Exp}(\beta)$.
This shows that $X^n_{(1)}$ is has an exponential
distribution with mean $1/(n\beta)$.
Conditioning on $X^n_{(1)}$, we get that $X^n_{(2)}-X^n_{(1)}$ is
again an exponential distribution equal to $X_{(1)}^{n-1}$ because of the memoryless
property of exponential distributions. We can repeat this argument to get the
density of $X^n_{(k+1)}-X^n_{(k)}$ for all $k$ up to $n-1$~\cite{Feller71}.

\section{Analysis}
\label{sec:partition}

In this section, we will show that our partition routine indeed constructs a
$(\beta,O(\frac{\log n}{\beta}))$ decomposition.
%In this section, we will describe our partition routine.
%Aside from being parallelizable, this routine can be viewed
%as a alternative to the repeated ball-growing arguments used
%in sequential algorithms.
For the purpose of this proof, we use a slightly different formulation of our
algorithm given in Algorithm \ref{alg:partition}. In this view, our algorithm
picks shifts $\delta_u$ for all vertices from independent exponential
distributions with parameter $\beta$, and then assigns each vertex to a piece
so that the shifted distances defined in (\ref{eqn:shiftDist}) to the center of
that piece is minimized.

\begin{algo}[ht]
\textsc{Partition}
\vspace{0.05cm}

\underline{Input:}
Undirected, unweighted graph $G = (V, E)$, parameter $\beta$ and
parameter $d$ indicating failure probability.

\underline{Output:} $(\beta, O(\log{n} / \beta))$ decomposition of $G$ with
probability at least $1 - n^{-d}$.

\begin{algorithmic}[1]
	\STATE{For each vertex $u$, pick $\delta_u$ independently from $\expdistr(\beta)$}
	\STATE{Compute $S_u$ by assigning each vertex $v$ to the vertex
		that minimizes $\distshift(u, v)$, breaking ties lexicographically}
	\RETURN{$\{ S_u \}$}
\end{algorithmic}

\caption{Partition Algorithm Using Exponentially Shifted Shortest Paths}
\label{alg:partition}

\end{algo}

We start by showing that the assignment process readily leads
to bounds on strong diameter.
Specifically, the strong diameter of $S_u$ can be measured using
distances from $u$ in the original graph.

\begin{lemma}
\label{lem:strongdiam}
If $v \in S_u$ and $v'$ is the last vertex on the shortest path from
$u$ to $v$, then $v' \in S_u$ as well.
\end{lemma}

\Proof
The proof is by contradiction, suppose $v'$ belongs to $S_{u'}$
for some $u' \neq u$.
The fact that $v'$ is the vertex before $v$ on the shortest path from
$u$ implies $\distshift(u, v) = \distshift(u, v') + 1$.
Also, as $v'$ is adjacent to $v$, we also have
$\distshift(u', v) \leq \distshift(u', v') + 1$.
Since $v'$ belongs to $S_{u'}$ instead of $S_{u}$,
we must have one of the following two cases:
\begin{enumerate}
\item $v'$ is strictly closer to $u'$ than $u$ in terms of shifted distance.
In this case we have $\distshift(u', v') < \distshift(u, v')$, which
when combined with the conditions above gives:
\begin{align*}
\distshift(u', v)
\leq & \distshift(u', v') + 1 \nonumber \\
< & \distshift(u, v') + 1 \nonumber \\
= & \distshift(u, v).
\end{align*}
So $v$ is strictly closer to $u'$ than $u$ as well, which implies
that $v$ should not be assigned to $S_u$.
\item The shifted distances are the same, and $u'$ is lexicographically
earlier than $u$.
Here a similar calculation gives $\distshift(u', v) \leq \distshift(u, v)$.
If the inequality holds strictly, we are back to the case above.
In case of equality, the assumption that $u'$ is lexicographically earlier
than $u$ means $v$ should not be in $S_u$ as well.
\end{enumerate}
\QED

Note that the second case is a zero probability event, and its
proof is included to account for roundings in implementations
that we will describe in Section \ref{sec:implementation}.

To bound the strong diameter of the pieces, it suffices
to bound the distance from a vertex to the center of the piece
that it is assigned to.
Since any vertex $v\in S_u$ could have been potentially included in $S_v$,
the shift value of the center $\delta_u$ serves as an upper bound on
the distance to any vertex in $S_u$.
Therefore, $\delta_{\max} = \max_{u} \delta_{u}$ serves as
an upper bound for the diameter of each piece.
Its expected value and concentration can be bounded as follows.

\begin{lemma}
\label{lem:maxshift}
The expected value of the maximum shift value is given by $H_n/\beta$ where
$H_n$ is the $n$th harmonic number. Furthermore, with high probability,
$\delta_u \leq O(\frac{\log{n}}{\beta})$ for all vertices $u$.
\end{lemma}

Our proof below proof closely following the presentation in
Chapter 1.6. of~\cite{Feller71}.

\Proof
The expected value can be found by summing over the differences
in order statistics given in Fact~\ref{fact:orderstatisticexp}.
\begin{align*}
  \expct{\max_{u\in V}\delta_u}&=\expct{X_{(n)}^n}
  \\&=\frac1\beta\sum_{i=1}^n\frac1n
  \\&=\frac{H_n}{\beta}.
\end{align*}

For the concentration bound, by the cumulative distribution function of the
exponential distribution
%given in Equation \ref{eq:cdf},
the probability of $\delta_u \geq \left( d+1 \right) \cdot \frac{\ln{n}}{\beta}$ is:
\begin{align*}
\exp \left( -(d+1) \cdot \beta \frac{\ln{n}}{\beta} \right)
=& \exp(-(d+1) \ln{n}) \nonumber \\
\leq & n^{-(d+1)}.
\end{align*}
Applying union bound over the $n$ vertices then gives the bound.
\QED

The other property that we need to show is that
few edges are between the pieces.
We do so by bounding the probability of two endpoints of an edge
being assigned to two different pieces.
In order to keep symmetry in this argument,
it is helpful to consider shifted distances from a vertex
to the midpoint of an edge.
This slight generalization can be formalized by replacing an edge
$uv$ with two length $1/2$ edges, $uw$ and $wv$.
We first show that an edge's end points can be in different pieces
only if there are two different vertices whose shifted shortest path
to its midpoint are within $1$ of the minimum.

\begin{lemma}
\label{lem:edgesplit}
Let $uv$ be an edge with midpoint $w$
such that when partitioned using shift values $\delta$,
$u \in S_{u'}$ and $v \in S_{v'}$.
Then both $\distshift(u', w)$ and $\distshift(v', w)$
are within $1$ of the minimum shifted distance to $w$.
\end{lemma}

\Proof
Let the pieces that contain $u$ and $v$ be $S_{u'}$ and $S_{v'}$
respectively ($u' \neq v'$).
Let the minimizer of $\distshift(x, w)$ be $w'$.
Since $w$ is distance $1/2$ from both $u$ and $v$, we have
\begin{align*}
\distshift(w', u), \distshift(w', v)
\leq & \distshift(w', w) + 1/2.
\end{align*}
Suppose $\distshift(u', w) >  \distshift(w', w) + 1$,
then we have:
\begin{align*}
\distshift(u', u)
\geq & \distshift(u', w) - 1/2 \nonumber\\
> & \distshift(w', w) + 1/2 \nonumber \\
\geq & \distshift(w', u),
\end{align*}
a contradiction with $u'$ being the minimizer of $\distshift(x, u)$.
The case with $v$ follows similarly.
\QED

An even more accurate characterization of this situation can be obtained
using the additional constraint that the shortest path from $w'$ to $w$
must go through one of $u$ or $v$.
%What Lemma \ref{lem:edgesplit} does is it allows us to
However, this lemma suffices for abstracting the situation further
to applying random decreases $\delta_1, \delta_2 \ldots \delta_n$
to a set of numbers $d_1 \ldots d_n$ corresponding to $\dist(x, w)$.
We now turn our attention to analyzing the probability of another shifted
value being close to the minimum when shifts are picked from the
exponential distribution.

The memoryless property of the exponential distribution gives
an intuitive way to bound this probability.
Instead of considering the vertices picking their shift values
independently, consider them as light bulbs with lifetime distributed according
to $\mathit{Exp}(\beta)$, and the $d_i$s indicate the time each light
bulb is being turned on. Then $\min_i d_i - \delta_i$ corresponds to the time
when the last light bulb burns out, and we want to bound the time between that
and the second last. In this setting, the memoryless property of exponentials
gives
that when the second to last light bulb fails, the behavior of the last light
bulb does not change and
its lifetime after that point still follows the same distribution.
Therefore, the probability that the difference between these two
is less than $c$ can be bounded using the
cumulative distribution function:
\begin{align*}
1 - \exp(-c \beta)
&\approx 1 - (1 - c \beta)\tag{When $c \beta$ is small} \\
&= c \beta.
\end{align*}
The only case that is not covered here is when the last light bulb has not been
turned on yet when the second last failed.
However, in that case this probability can only be less.
We give a rigorous version of this intuitive proof below.
An algebraic proof using the definition of the exponential distribution
can be found in Appendix~\ref{sec:altproof}.

\begin{lemma}
\label{lem:closeprob}
Let $d_1 \leq \ldots \leq d_n$ be arbitrary values and $\delta_1
\ldots \delta_n$ be independent random variables picked from
$\expdistr(\beta)$. Then the probability that between the smallest
and the second smallest values of $d_i-\delta_i$ are within $c$
of each other is at most $O(\beta c)$.
%is distributed according to $\mathit{Exp}(\beta)$.
\end{lemma}

\Proof
It is more convenient to consider the differences between
the largest and second largest of the negations of the shifted
values, $-(d_i - \delta_i)$.
%As $d_i$s can be any values, we will use $d'_i$ to denote $-d_i$.
%For notational convenience, we prove an equivalent statement where we consider
%positives shifts $d_i+\delta_i$ and give probabilistic bounds on the difference
%between the largest value and the second largest value.
Let $d'_i$ denote $-d_i$, by the assumption of $d_1 \leq \ldots \leq d_n$
we have $d_1' \geq \ldots \geq d_n'$.
Define $X_i=d_i'+\delta_i-d'_1$ and let $X_{(i)}$ denote the $i$th
order statistic of $X_1,\dots,X_n$, we would like to show that
\begin{align*}
  \prob{X_{(n)}-X_{(n-1)}>c}\ge \exp(\beta c).
\end{align*}

Since $X_i$s are independent, the memoryless property of exponential
distributions gives that when conditioned on $X_i \ge 0$, $X_i$
still follows an exponential distribution with mean $1/\beta$.
For all subsets $S\subseteq\{1 \ldots n\}$, let
$\mathcal{E}_S$ denote the event that for all $i\in S$, $X_i\ge0$,
and for all $i\notin S$, $X_i<0$.
By the law of total probability, we have
\begin{align*}
&  \prob{X_{(n)}-X_{(n-1)}>c}\\
&  =
  \sum_S
  \prob{X_{(n)}-X_{(n-1)}>c\mid \mathcal{E}_S}
  \prob{\mathcal{E}_S}.
\end{align*}
Since $X_n = \delta_n \geq 0$, $\prob{\mathcal{E}_S}=0$
when $S=\emptyset$ or $S\not\ni n$.
The only other case with $|S| = 1$ is when $S=\{n\}$.
Here we have $\prob{X_{1}>c} \geq 1 - \exp(\beta c)$.
Combining this with $X_{(n)} \geq X_1$ and $X_{(n - 1)} < 0$ gives
a probability of at least $\exp(\beta c)$.

It remains to consider the case where $|S| \geq 2$.
In this case both $X_{(n)}$ and $X_{(n - 1)}$ are from elements in $S$,
so it suffices to consider the $X_i$s given by $i\in S$.
These $|S|$ variables are distributed the same as $|S|$ independent
random variables following $\expdistr(\beta)$.
Therefore by the distribution of order statistics given in
Fact~\ref{fact:orderstatisticexp} we have:
\begin{align*}
  \prob{X_{(n)}-X_{(n-1)}>c \mid \mathcal{E}_{S}, |S| \geq 2}=\exp(-\beta c).
\end{align*}
This means for any $S$, we have
\begin{align*}
\prob{X_{(n)}-X_{(n-1)}>c \mid \mathcal{E}_S} \prob{\mathcal{E}_S}
& \geq \exp\left( -\beta c\right) \prob{\mathcal{E}_S}.
\end{align*}
Summing over all $S$ and using the fact that
$\sum_{S}\prob{\mathcal{E}_S} = 1$ gives that
\begin{align*}
\prob{X_{(n)}-X_{(n-1)}>c}\ge\exp\left(-\beta c\right),
\end{align*}
or equivalently
\begin{align*}
\prob{X_{(n)}-X_{(n-1)}\le c}\le1-\exp\left(-\beta c\right)<\beta c.
\end{align*}
\QED

Using this Lemma with $c = 1$ and applying linearity of
expectation gives the bound on the number of edges
between pieces.

\begin{corollary}
\label{cor:edgescut}
The probability of an edge $e = uv$ having $u$
and $v$ in different pieces is bounded by $O(\beta)$,
and the expected number of edges between pieces is $O(\beta m)$.
\end{corollary}

\section{Implementation and\\Parallelization}
\label{sec:implementation}

Our partition routine as described in Algorithm \ref{alg:partition}
requires computing $\distshift(u, v)$ for all pairs of vertices $u$ and $v$.
Standard modifications allow us to simplify it to the form
shown in Algorithm~\ref{alg:parallelPartition}, which
computes BFS involving small integer distances.

The first observation is that the $-\delta_u$ shift at vertex $u$
can be simulated by introducing a super source $s$ with distance
$-\delta_u$ to each vertex $u$.
Then if we compute single source shortest path from $s$ to all vertices,
the component that $v$ belongs to is given by the first vertex on
the shortest path from $s$ to it.
Two more observations are needed to transform this shortest
path setup to a BFS.
First, the negative lengths on edges leaving $s$ can be fixed by
adding $\delta_{\max} = \max_{u} \delta_u$ to all these weights.
Second, note that the only edges with non-integral lengths
are the ones leaving $s$.
In this shortest path algorithm, the only time that we need to
examine the non-integer parts of lengths is when we compare
two distances whose integer parts are tied.
So the fractional parts can be viewed as tie-breakers for equal
integer distances, and all distances with the same integer part
can be processed in parallel.
We'll show below that these tie breakers can also be replaced
by a random permutation of integers.

Therefore, the algorithm is equivalent to computing shortest path
when all edge lengths are integer, with an extra tie breaking rule
for comparing distances.
In order to use unweighted BFS, it remains to handle the edges
with non-unit lengths leaving $s$, and we do so by processing
the those edges in a delayed manner.
An edge from $s$ to $v$ only causes $v$ to be added to the BFS
queue when the frontier of the search has reached a distance
larger than the length of that edge and $v$ has not been visited yet.
So it suffices to check all vertices $v$ with a length $L$ edge to $s$
when the BFS frontier moves to distance $L$, and add the
unvisited ones to that level.

The exact cost of running a parallel breadth first search depends
on the model of parallelism.
There has been much practical work on such routines when the
graph has small diameter \cite{LeisersonS10, BeamerAP12, ShunB13}.
For simplicity we will use the $O(\Delta \log{n})$ depth
and $O(m)$ work bound in the PRAM model given in~\cite{KleinS97}.
Here $\Delta$ is the maximum distance that we run the BFS to,
and can be bounded by $O ( \frac{\log{n}}{\beta} )$.
This allows us to prove our main claim about the performance
of \textsc{Partition}.

\Proofof{Theorem \ref{thm:partition}}
Consider running \textsc{Partition} using the BFS based
implementation described above, and repeating until we have an
$(\beta, O(\frac{\log{n}}{\beta}))$ partition.
Since the $\delta_u$s are generated independently, they can be
computed in $O(n)$ work and $O(1)$ depth in parallel.
The rest of the running time comes from assigning vertices to
pieces using shifted shortest path.
As the maximum distance from a vertex to the center of its
piece is $O(\frac{\log{n}}{\beta})$ (or we could stop the algorithm at
this point), this BFS can be done in $O(m)$
work and $O(\frac{\log^2{n}}{\beta})$ depth using parallel BFS algorithms.
The resulting decomposition can also be verified in $O(\log{n})$
depth and $O(m)$ time.

It remains to bound the success probability of each iteration.
Lemma \ref{lem:strongdiam} gives that the shortest path
from $u$ to any $v \in S_u$ is contained in $S_u$,
so $\max_{v \in S_u} \dist(u, v)$ is an upper bound
for the strong diameter of each subset.
For each vertex $v \in S_u$, since $\distshift(v, v) =
d(v, v)  - \delta_v \leq 0$ is a candidate,
$\distshift(u, v) \leq -\delta_v$.
Lemma \ref{lem:maxshift} then allows us to bound this value by
$O(\log{n} / \beta)$ with high probability.
The expected number of edges between pieces follows from
Corollary \ref{cor:edgescut}, so with constant probability we
meet both requirements of a
$(\beta, O(\frac{\log{n}}{\beta}))$ partition.
Therefore, we are expected to iterate a constant number of times,
giving the expected depth and work bounds.
\QED

One practical aspect worth noting is that the fractional parts
of the $\delta$ values can be viewed as a lexicographical
ordering upon all vertices which are used for tie breaking.
This is where the tie breaking rule specified in
Section~\ref{sec:partition} may be of use.
As the exponential distribution is memoryless and the shifts are
generated independently, the fractional parts can also be emulated
by directly generating a random permutation of the vertices.
This view is perhaps closer to the use of random permutations
in the optimal tree-metric embedding algorithm
\cite{FakcharoenpholRT03}.

Similar ideas may also be used in practice instead of computing
$\delta_u$.
Although generating random variables from such distributions have
been studied extensively \cite{KnuthBook2}, avoiding these routines
might further reduce the cost of this stage of the algorithm.
One possibility is to generate a random permutation of the vertices,
and assign the shift values based on positions in the permutation.
We believe that the slight changes in distributions could be
accounted for using a more intricate analysis, but might be
more easily studied empirically.

\section{Conclusion / Remarks}
\label{sec:conclusion}

We showed a simple parallel algorithm for computing low diameter
decompositions of undirected unweighted graphs.
Given a graph $G$ with $n$ vertices and $m$ edges along with
any parameter $\beta$, it returns a $(\beta, O(\frac{\log{n}}{\beta}))$
decomposition in $O(\frac{\log^2{n}}{\beta})$ depth and $O(m)$ work.
This routine can be used in place of \textsc{Partition} from
\cite{BlellochGKMPT11} to give a faster algorithm for solving
SDD linear systems.
It also represents a different view of ball growing, which is at the
core of the best sequential low stretch spanning tree algorithms
\cite{ElkinEST08,AbrahamBN08,AbrahamN12}.

We believe that our approach may lead to a variety of improvements
in algorithms that use ball growing or low diameter decompositions.
Many of these applications take place in the weighted setting,
and rely on additional clustering-based properties.
As a result, many of them are perhaps better examined on a per-application basis.
The analysis of the partition routine from Section~\ref{sec:partition}
can be readily extended to the weighted case.
However, the depth of the algorithm is harder to control since hop
count is no longer closely related to diameter.
We believe obtaining similar parallel guarantees in
the weighted setting, as well as showing clustering-based
properties are interesting directions for future work.

% Please use sec:abc thm:abc and lem:abc for all labels.  Each section (and
% subsection), theorem and lemma must have a meaningful label.

%\vfill                          % balance columns
%\begin{spacing}{0.7}
%  \begin{small}
\bibliographystyle{abbrv}
%\bibliography{../../../Bibtex/ref,../../../Bibtex/rpeng,../../../Bibtex/vision,../../../Bibtex/vision2}
%\bibliography{../../../Bibtex/ref,../../../Bibtex/miller,../../../Bibtex/vision,../../../Bibtex/vision2,../../../Bibtex/rpeng,../../../Bibtex/graph}
%  \end{small}
%\end{spacing}

\begin{appendix}

\section{Alternate Proof of Key Partition Lemma}
\label{sec:altproof}

We give alternate, more formulaic proofs of
Lemma~\ref{lem:maxshift}, which bounds
the difference between the shortest and second
shortest shifted distance to any point in the graph.

\Proofof{Lemma~\ref{lem:closeprob}}

Let $\mathcal{E}$ denote the number of indices $i$ such that:
\begin{align*}
	d_i - \delta_i \leq d_j - \delta_{j} + c ~\forall j.
\end{align*}
For each vertex $i$, let $\mathcal{E}_i$ be an indicator variable
for the event that:
\begin{align*}
	d_i - \delta_i \leq d_j - \delta_{j} + c ~\forall j.
\end{align*}
We will integrate over the value of $t = d_i - \delta_i$.
For a fixed value of $t$, $\mathcal{E}_i$ occurs if and only if
$\delta_{j} \leq d_j - t + c$ for each $j$.
As the shift values are picked independently, we can multiply
the cumulative distribution functions for $\expdistr(\beta)$
and get:
\begin{align*}
& \prob{\mathcal{E}_i}\\
= & \int_{t = -\infty}^{\infty} f_{Exp}(d_i - t, \beta) \prod_{j \neq i}
	F_{Exp}(d_j - t + c, \beta).
\end{align*}
When $t > d_1 + c$, $d_1 - t + c < 0$ and
$f_{Exp}(d_1 - t, \beta) = F_{Exp}(d_1 - t + c, \beta) = 0$.
So it suffices to evaluate this integral up to $t = d_1 + c$.
Also, we may use $\exp(-\beta x)$ as an upper bound as
$f_{Exp}(x , \beta)$, and arrive at:
\begin{align*}
& \prob{\mathcal{E}_i}\\
& \leq \int_{t = -\infty}^{d_1 + c} \beta \exp(-\beta(d_i - t))
	\prod_{j \neq i}	F_{Exp}(d_j - t + c) \nonumber\\
& \leq  \int_{t = -\infty}^{d_1 + c} \beta \exp(-\beta(d_i - t))
	\prod_{j \neq i} \left( 1 - \exp(-\beta(d_j - t + c))) \right).
\end{align*}
We now bound $\expct{\mathcal{E}} = \expct{\sum_{i} \mathcal{E}_i}$.
By linearity of expectation we have:
\begin{align*}
\expct{\sum_{i} \mathcal{E}_i}
\leq & \sum_{i}  \int_{t = -\infty}^{d_1 + c} \beta
		\exp\left( -\beta(d_i - t) \right)
\nonumber \\ & \qquad
			\prod_{j \neq i} \left( 1 - \exp(-\beta(d_j - (t - c))) \right) \nonumber \\
= & \exp(\beta c) \int_{t = -\infty}^{d_1 + c}  \beta
		\sum_{i} \exp \left( -\beta(d_i - t + c) \right)
\nonumber \\ & \qquad
			\prod_{j \neq i} \left( 1 - \exp(-\beta(d_j - t + c) ) \right).
\end{align*}
Observe that the expression being integrated is the derivative w.r.t. $t$ of:
\begin{align*}
- \prod_{i} \left( 1 - \exp( - \beta(d_i - t + c)) \right).
\end{align*}
Therefore we get:
\begin{align*}
\expct{\mathcal{E}}
\leq & - \exp(\beta c)
 \left. \prod_{i} \left( 1 - \exp( - \beta(d_i - t + c))  \right) \right|
	_{t = -\infty}^{t = d_1 + c}
\end{align*}
When $t \rightarrow -\infty$, $-\beta(d_i - t + c) \rightarrow -\infty$.
Therefore $\exp( - \beta(d_i - t + c)) \rightarrow 0$,
and the overall product tends to $-\exp(\beta c)$.

When $t = d_1 + c$, we have:
\begin{align*}
& - \exp(\beta c)\prod_{i} \left( 1 - \exp( - \beta(d_i - (d_1 + c) + c) )  \right) \nonumber \\
= & - \exp(\beta c)\prod_{i} \left( 1 - \exp(-\beta (d_i - d_1) )  \right) \nonumber \\
\leq & - \exp(\beta c) \prod_{i} \left( 1 - \exp(0)  \right) = 0
	\tag{Since $d_i \geq d_1$}
\end{align*}
Combining these two gives $\expct{\mathcal{E}} \leq \exp(\beta c)$.

By Markov's inequality the probability of there being another vertex
being within $c$ of the minimum is at most $\exp(\beta c)  - 1 \leq O(\beta c)$
for $c=1$.
\QED

\end{appendix}

%\begin{appendix}
%\end{appendix}

\end{document}